\documentstyle[preprint,aps,prbbib]{revtex}
\draft
\begin{document}
\title{
Role of structural relaxations, chemical substitutions
and polarization fields  on the potential
line--up in [0001] wurtzite GaN/Al systems}

\author{S. Picozzi, G. Profeta and A. Continenza}
\address{Istituto Nazionale di Fisica della Materia (INFM)\\
Dipartimento di Fisica,
Universit\`a degli Studi di L'Aquila, 67010 Coppito (L'Aquila), Italy }
\author{S. Massidda}
\address{Istituto Nazionale di Fisica della Materia (INFM)  \\
Dipartimento di Scienze Fisiche \\ Universit\`a degli
Studi di Cagliari,  09124 Cagliari, Italy \\ and}
\author{A. J. Freeman}
\address{Department of Physics and Astronomy and Materials Research Center\\
Northwestern University, Evanston, IL 60208 (U.S.A.)\\}

\maketitle

\begin{abstract}
First--principles full--potential linearized augmented plane wave (FLAPW)
 calculations are performed to
clarify the role of the interface geometry on 
piezoelectric fields and on  potential line--ups
at the [0001]-wurtzite and  [111]-zincblende GaN/Al junctions. 
The electric fields (polarity and
magnitude) are found to be strongly affected by
atomic relaxations in the interface region. A procedure is tested
to evaluate the
Schottky barrier in the presence of electric fields and used to
show that their effect 
is quite small
(a few tenths of an eV).
These calculations 
assess the rectifying behaviour of the GaN/Al contact, giving very 
good agreement
with experimental values for the barrier. Stimulated by the 
complexity of the
problem, we disentangle  chemical and structural effects on the
relevant  
properties (such as the potential discontinuity and electric fields) by
studying  auxiliary unrelaxed nitride/metal systems.
Focusing on simple electronegativity arguments, 
we outline the leading mechanisms 
that result in the final values of the
electric fields and Schottky barriers in these ideal interfaces.
Finally, the transitivity rule in the presence of two inequivalent junctions
is proved to give reliable results.
\end{abstract}

\section{Introduction}

Wide-bandgap electronic devices are expected to play an important role in
the next generation of high--power and high--temperature applications.
In particular, extensive work has been carried out in recent years on
many popular semiconductors of this type -
GaN, AlN and SiC\cite{review,binggeli}. It is clear,
however, that many efforts, theoretical as well as experimental, are
still needed to bring the emergent device technology 
to maturity. Within this context, the GaN/metal interface represents
one of the most well--studied
 topics \cite{review}, since it is not yet clear  which are the basic
mechanisms leading to the observed behavior - ohmic vs rectifying  - of the
final barrier.
A key property of the nitrides is the presence of large 
spontaneous and piezoelectric fields, which have to be considered
whenever these compounds are used as basic constituents of  technological devices
\cite{fabiopol}.
Within the framework of {\em ab--initio} simulations, extensive 
work has been carried out
for nitride/metal contacts, but 
the GaN structure explored so far has always been 
zincblende  with  [001]  as the ordering direction \cite{slv}.
In contrast, however,   
 most of the experimental work in this field is
 focused on the hexagonal wurtzite structure as the stable phase of GaN.
\cite{ambacher,bermudez}.

In this work, we present  results obtained from  first-principles
calculations using
the all--electron full--potential linearized augmented plane
wave (FLAPW)\cite{FLAPW} method for [0001] w-GaN/Al and [111] z-GaN/Al interfaces.
These systems can be directly compared,
since [111]--ordered zincblende and [0001]--ordered wurtzite
have exactly the same coordination up to the third nearest neighbor shell and
the interface geometry (in terms of number and direction
of bonds between different atomic species at
the junction) is exactly the same.

\section{Computational  details}

The calculations were performed using one of the most accurate and highly
precise first principles density--functional  based
methods, namely the all-electron  FLAPW
code \cite{FLAPW}. 
The core levels are treated fully relativistically, whereas the
valence electrons are calculated semi--relativistically, {\em i.e.} without 
spin-orbit coupling. The Ga 3$d$ states were treated as valence electrons, 
in order to fully allow their characteristic hybridization
 with the N 1$s$ states. The exchange--correlation potential is
treated within the local density approximation using the Hedin--Lundqvist
parametrization\cite{hl}. The FLAPW code allows calculation of
 total energy and atomic--forces, so that
the structural optimization of the atomic positions is achieved by first
principles. The muffin--tin spheres were chosen as $R_{MT}^{Ga}$ = 2.0 a.u.,
$R_{MT}^{Al}$ = 2.0 a.u. and $R_{MT}^{N}$ = 1.6 a.u. and the expansion in
spherical harmonics in these regions was performed up to $l \leq$ 8; in the
interstitial part, a cut--off for the wave functions $k_{max}$ = 4.0
a.u.$^{-1}$ was used.
The Brillouin zone sampling was done according to the Monkhorst--Pack scheme
\cite{mp},
using 10 {\bf k} points in the irreducible part of the  zone.

\section{Structural properties and energetics}

The unit cell employed  contains 6 layers of GaN ({\em i.e.} 7
N and 6 Ga atoms) and 6 layers of Al, for a total of 19 atoms per cell.
Tests performed with varying cell dimensions have shown that this particular choice
of number of layers for each material ({\em i.e.} semiconductor and metal)
is sufficient to recover the proper bulk conditions away from the
interface: this represents a major requirement when dealing with potential
line--up problems with the supercell approach.

The equilibrium atomic-force and total--energy 
optimized w-GaN lattice constants are $a$ = 3.16
 \AA $\:$ and $c/a$ = 1.624,
with an internal parameter $u$ = 0.377. The agreement with the experimental
values\cite{lb} $a$ = 3.16, $c/a$ = 1.62 and $u$ = 0.377 is excellent, 
as is usual for FLAPW structural results obtained for III-V semiconductors.

In the following, we will discuss in detail all the different structures
studied, focusing on their differences and similarities.

\subsection{Relaxed [0001] w-GaN/Al and [111] z-GaN/Al}

As far as the structural configuration is concerned, we remark
that optimization of atomic positions is very delicate in the GaN/Al system.
In fact, as shown in the  z-GaN[001]/Al case, the Schottky barrier is very
sensitive to the interface geometry, namely the interplanar distances in
proximity to the junction\cite{slv}. Due to the 
polarity of the interface in the
 systems of interest here, we may expect an
even stronger effect of the structural configuration. In fact, in both
    N--terminated   w-GaN[0001]/Al and   z-GaN[111]/Al, the
two interfaces present in the unit cell are inequivalent; as clearly
shown  in Fig. \ref{cell}, the interface N is bonded with 3 Ga and 1 Al on one
side (A--type with the  ``long" Al-N bond parallel to the growth axis)
and with 3 Al and 1 Ga on the other side (B--type with the  ``short" 
Al-N bond parallel to the growth axis, leading to an interplanar distance
which is ideally one third of the bond length)\cite{nota_cl}. 

All the atomic positions were allowed to relax; the relevant
interface bond lengths in the optimized geometries are reported in
Table \ref{distan}.  First of all, we note that the differences
between the zincblende and wurtzite systems are negligible (less than 0.1 $\%$
of the bond lengths) and this confirms the similarity of the z-GaN[111]/Al
and w-GaN[0001]/Al. The  Ga-N   distances can be compared with
 the w-GaN bulk bond length of 1.92 \AA $\:$ and
 are  found to deviate only by 1 $\%$ with respect to
  the equilibrium  Ga-N bond length,  showing that 
the  relaxations  in the semiconductor side are rather small. 
As expected,
only the {\em interface} atoms in the semiconductor
side  deviate slightly from the bulk w-GaN distances;  already in the second
layer, the first--nearest--neighbor bond lengths typical of bulk w-GaN are
 recovered (within 0.3 $\%$). On the other hand, the Al atoms   in the
metallic layer deviate from the typical interplanar Al-Al
distance  for  [111] ordered fcc-Al strained on
[0001] w-GaN ($z_{Al-Al}$ = 2.046 $\AA$) by as much as 10 $\%$. However,
the positions of the Al atoms far from the junction  are not expected
to play a major role in the evaluation of the Schottky barrier heights
(see discussion below). Moreover, we find some 
   major differences in the relaxations at
the two inequivalent interfaces, which  are mainly
 related to the Al-N bond lengths. At the
B interface, the ``short" Al-N bond length ($d_{Al-N}^A$ = 1.93 \AA) is more or
less unaltered with respect to the bulk GaN bond length (by
$<$ 0.4 $\%$), so
that  Al perfectly replaces the Ga cation in a
bulk GaN. On the other hand, at the A interface the ``long" Al-N bond length
 ($d_{Al-N}^B$ = 1.89 \AA)
 is smaller by
2.2 $\%$ than the bulk  GaN bond length and much closer to the
  bulk w-AlN bond length \cite{lb} ($d_{Al-N}^{w-bulk}$ = 1.87 \AA ).
Therefore, the tendency of Al and N atoms to get closer is higher (lower)
when they are aligned (not--aligned) along the growth direction.

\subsection{Model GaN/Al systems}

In  order to further investigate the {\em effect of the relaxations}, we
also considered: 

\begin{itemize} 

\item  An ``ideal" ({\em i.e.} unstrained)
system (denoted as w-GaN/Al$^{id}$) obtained starting from [0001] ordered bulk w-GaN,
where the anions in the second
half of the cell have been removed and the Ga cations 
substituted with Al.  This gives rise
to a  semiconductor/metal interface between w-GaN and Al with all the Ga-N and
Al-N distances equal to those in w-GaN, while the Al-Al distances match those 
between 
cations in w-GaN. 
\item A ``partially relaxed" system (denoted  as
w-GaN/Al$^{pr}$) where all the interface and subinterface interplanar distances
are relaxed to those in the fully relaxed system 
and the Al-Al distances in the metal side are forced to be equal.
This system will be helpful to understand the effects of displacing Al atoms far from the
junction in terms of stability, SBH and
electric fields.

\end{itemize}

We focus first on the stability of the  different systems. We show in
Table \ref{energy} the total energies  per atomic species  of the systems
referred to the most stable interface ({\em i.e.} w-GaN/Al$^{rel}$)
taken as reference: these quantities represent 
the energy gain 
per atom with respect to the fully relaxed system.
We observe that the effect of the
relaxations in the interface region is quite high (about 50 meV/atom in going
from  w-GaN/Al$^{id}$ to  w-GaN/Al$^{pr}$), even if
the deviations in terms of bulk
bond lengths are less than 2 $\%$. On the other hand, the
total energy is very stable when the Al atoms in the bulk metal region
are relaxed: 
less than 5 meV/atom.
As expected, it is very important to accurately optimize the
interface geometry, while  changes of the atomic positions within the bulk
metallic region are far less energetically
expensive.
Also, in agreement with previous results\cite{az},  we found that
 zincblende 
GaN is not favored compared to the   wurtzite phase (but only by at most 10
meV/atom)\cite{consist}.

\subsection{Additional model systems}

\begin{itemize} 

\item {\em Unrelaxed z-XN/Y,
(X, Y = Ga, Al)}

In order to separate {\em chemical} and {\em structural} contributions to the
SBH, 
we considered some unrelaxed [111] ordered
(z-XN/Y)$^{id}$ systems, where X,Y = Ga, Al.
In  these ideal systems, all the N-X (X = Ga, Al) bond
lengths are equal, leading to    more ``symmetric"  junctions (compared to the
relaxed systems). Moreover, let us recall that the zincblende undistorted
structure does not show any
spontaneous polarization fields because of symmetry.
Now, these model nitride/metal interfaces are not meant to simulate  realistic
systems; in fact,  $\alpha$-Ga being the Ga stable structure, the
interface relaxations and coordinations in the XN/Ga junctions
would be completely different from
those considered here. Nevertheless, we will take
these ideal systems  
as 
reference structures: indeed,
in these junctions 
the atomic positions are frozen and set equal; 
only the chemical species
occupying the atomic sites differ, thus allowing us to separate 
chemical from structural effects. 

Insights into the chemical bonding at
 the nitride/metal systems can be gained by focusing 
on the adhesion energy $E_{ad}$ of these unrelaxed
systems. 
In order to estimate the gain in energy when depositing  Al
or  Ga  on a nitride surface, we calculate the difference between the
adhesion energies \cite{adhes} in the (GaN/Ga)$^{id}$ and (GaN/Al)$^{id}$ junctions;
$E_{ad} (GaN/Al)^{id} -  E_{ad} (GaN/Ga)^{id}$ =  1.04 eV
and $E_{ad} (AlN/Al)^{id} -  E_{ad} (AlN/Ga)^{id}$ = 1.06 eV.
The  positive sign of these energy values indicates the larger energy gain 
that favours deposition of 
Al versus Ga
on the N-terminated nitride surface; this is in agreement with the
more ionic character of the Al-N bond 
compared to  Ga-N. Moreover, the
similarity of the two values confirms that the gain in adhesion 
energy is similar for the two nitride surfaces.

\item {\em Unrelaxed
 z-XN/Y/X and z-YN/X/Y
(X, Y = Ga, Al)}

Finally, with the only purpose 
 to {\em investigate the role of a single metallic 
interlayer on the final line-up and
 electric fields}, we also focused on two systems,
(z-AlN/Ga$^{int}$/Al)$^{id}$ and (z-GaN/Al$^{int}$/Ga)$^{id}$.
 These systems are
obtained starting from the ideal (z-XN/X)$^{id}$ interfaces and substituting the
first X atom on the metallic side for the Y cation ({\em i.e.} we start from the
(z-AlN/Al)$^{id}$ systems and substitute the first Al layer with Ga, ending up with
the
(z-AlN/Ga$^{int}$/Al)$^{id}$ system.)
\end{itemize}

\section{Electronic properties}

Let us investigate the two inequivalent interfaces (A and B in Fig.\ref{cell}) 
in terms of their density
of states of the wurtzite based systems (the z-GaN/Al is very similar and
therefore not discussed). In Fig. \ref{fig_dos}, we show the density of states
(PDOS) projected on the interface Ga, N and Al 
atomic sites in the relaxed w-GaN/Al 
junction for the A--type  (panels (a), (b) and (c),
respectively)  and for the B type  interface (panels (d), (e) and (f)). Both
Fig. \ref{fig_dos} (a) and (d) show that  the sub-interface Ga
atoms in both A and B--type junctions have a PDOS very similar to that of 
the Ga
atom in bulk w-GaN, except for the appreciable DOS due to the
metal induced gap states (MIGS). 
On the other hand, the interface N atoms clearly differ in the A and
B--type junctions; note, in particular, that the  feature at -8 eV 
in panel (b) , mainly due to cation
$s$ states, is shifted towards lower binding
energies (about -7 eV  in panel (e)). A more careful investigation shows that
many of the features in the N PDOS of the A--type
 junction are common to   
bulk w-GaN (dashed line in panel (b)),
while the N PDOS in the B--type interface is more similar to the one in bulk
w-AlN (dashed line in panel (e)).
Furthermore,  the Al interface DOS  
is remarkably different in the two inequivalent junctions: in terms of the
PDOS,  this atom resembles the cation in bulk w-AlN in type B (dashed line
in panel (f)), whereas it shows a
free--electron--like behaviour, close to the one in bulk fcc-Al (dashed line in
panel (c)) in the A-type interface.

 The metallicity of the Al atom is confirmed by the
higher DOS in the band-gap region ({\em i.e.} from about -2 eV
to the Fermi level, $E_F$) of the Al in the
A--type  compared to that in the
B--type junction. A simple  rationale can be found by considering the
geometry of the two inequivalent interfaces shown in Fig.1; 
 in the A type the four N $sp^3$ orbitals point
towards three Ga and one Al, whereas the situation is reversed ({\em i.e.}
three Al and one Ga) in the B type junction. Thus, we can reasonably expect
the  N atom to be similar to  bulk w-GaN (w-AlN) in the A (B) interface.
Moreover, note that the higher DOS at  $E_F$ of the
A-type Al atom (panel (c)) suggests a more effective screening than that at
the B interface (panel (f)). 

Finally, the
  bond length does not greatly affect   the DOS: in fact,
the  PDOS (not shown) in the ``frozen" reference structure (w-GaN/Al)$^{id}$,
having all the AlN and GaN distances equal to those in bulk w-GaN, shows
features that are
very similar to the PDOS of the relaxed system (where every cation-N
distance is different). Therefore, the electronic 
behaviour is dictated
by the number and direction of bonds established between different atoms
rather than by the distance between the atomic species participating in the
bond.

\section{Electric fields}
\label{elecfield}

The inequivalency of the A and B interfaces, in terms of different geometries
and bond lengths, gives rise to electric fields; in addition, in the
wurtzite--based systems, we  expect  the presence of intrinsic
polarization fields, which vanish by symmetry in the zincblende--based
junctions. 
In Fig. \ref{cori}, we show the N $1s$  core
level binding energies (in eV) in the wurtzite based systems for the ideal,
partially and fully relaxed systems; the binding energies have been
 shifted arbitrarily so as to let the central core level  coincide with the
zero. The slope of the excellent linear fits through the $1s$ binding energies gives
directly the electric fields, and their magnitudes (in V/nm) 
are reported in Table \ref{table_cori} for all the systems
examined. The interface N
atoms have been excluded from the fit, since, due to charge
rearrangement in proximity to the junction, they 
are subject to additional local effects.

We expect 
that the polarization charge  giving rise to the electric field
is strongly affected by the interface geometry: in fact,  the
polarity of the field  changes in going from ideal to relaxed systems and,
in particular, the field is directed from the A to the B interface  in the
relaxed systems, whereas the situation is reversed in the ideal case.
 
 The mechanisms giving rise to the electric field are
very complicated: the interplay of boundary conditions, charge redistribution
at the inequivalent interfaces and screening effects suggest that the value
and even the polarity of the field cannot be determined within simple
electrostatic or electronegativity arguments. In this context, 
{\em ab--initio} simulations  are the only way 
to take into account   microscopic details of  charge
rearrangement 
and  boundary conditions, giving a correct description of the
overall electrostatics.
It can be argued that the almost
negligible electric field present in the unrelaxed interface 
is due to the tendency of
Al to  screen the electric field: 
this is undoubtedly true for every
good metal. However,   the large total energy difference between the
ideal and relaxed interfaces in Table \ref{energy} suggests that
 the electrostatic
energy accumulated in the semiconductor side by the field 
is negligible by far 
compared to other total energy terms related to structural relaxations.
Furthermore, the differences between
the values of the electric fields in the zincblende and wurtzite systems are
        quite small so that the presence of intrinsic polarization fields is
not playing a significant role in the final determination of the
potential line--up (see below).

Our results for the electric fields in the ideal systems are shown in Fig.
\ref{gianni}. The polarity of the field is the same in all the ideal systems
(the electric field going from the B to the A-type interface), but the
magnitude shows large variations. 
First note that 
a naive picture for the  
(z-GaN/Al) interface (see Fig. \ref{cori})
 would suggest that more symmetric interfaces ({\em i.e.}
with similar anion-cation bond lengths at the two interfaces)
would exhibit smaller fields. 
 However, as shown in Fig. \ref{gianni},  this is not the case. Consider, for example,
the (z-GaN/Ga)$^{id}$  ideal interface whose atomic
positions and species are exactly equivalent to the
 (z-GaN/Al)$^{id}$, with Ga replacing Al in the metal side: here the
electric field has the same order of magnitude as in the relaxed systems 
(see Fig. \ref{cori}),
but with reversed polarity.
This  confirms the complexity of the problem and of the interplay of strain and
chemical effects. However, we can give a rationale for some of the observed
behavior. For example, we note that in all the systems with Al (Ga) at the
 interface, the electric field is negligible (quite large); 
therefore, it seems that the Al atom favors charge rearrangement 
(whether it comes from GaN or AlN)  in such a way as to screen the electric field
much more efficiently than  Ga.  
As for the comparison between the two different nitrides, we expect on the basis
of electronegativity and ionicity arguments a larger interface bond 
charge in AlN
than  in GaN: this is confirmed by most of the results
shown in Fig. \ref{gianni}. 

In order to clarify the role of the interface
metal layer, we plot in Fig. \ref{sandro} the difference between the
macroscopic planar average of the valence charge densities for the  
(z-GaN/Ga)$^{id}$ and (z-GaN/Al$^{int}/$Ga)$^{id}$ systems as a function of the
growth direction.
The only difference between these two systems 
is the metal atom bonded with N;
however, the resulting electric fields are very different
 (see Fig. \ref{gianni}). This shows that the major role in establishing 
the final field is played by
the interface metal  atom rather than by the screening properties of the 
metal side of the junction. In other terms, the electric field, at fixed
periodic boundary conditions, is determined by ``interface" effects rather than
by ``bulk" properties of the constituents. 
Moreover, Fig. \ref{sandro} demonstrates the different 
electronegativity between Al and Ga: more charge is localized on the
interface N atom when it is bonded to Al than to Ga; 
also, the charge difference integral (dashed line in
Fig. \ref{sandro}) shows that
this effect is more pronounced in the A-type
interface, where the metal-N bond is  parallel to the growth axis.
This is in agreement with what was pointed out for the relaxed 
(GaN/Al)$^{rel}$ systems: due to the particular interface geometry,
the A-type interface favors charge transfer from Al to N 
with respect to the B-type junction.    

Finally, we observe that the effect of the Ga $d$ electrons on the electric
fields is absolutely
negligible; in fact, if we consider the (z-GaN/Ga)$^{id}$ interface and treat 
the Ga $d$ states as {\em i}) valence or as {\em ii}) core electrons, we obtain
exactly the same resulting electric field.

   \section{Schottky barrier
heights}
\label{sbh_sec}
Let us now discuss the most technologically important property,
 the Schottky barrier height and its relation to the piezoelectric
fields. Usually, within all--electron methods the
SBH is evaluated following a procedure based on core levels taken as reference
energies \cite{min}. The final SBH is therefore expressed as:
\begin{equation}
\Phi = \Delta b + \Delta E_b
\label{sbh}
\end{equation}
where  $\Delta b$ is the difference between the 1$s$ Ga and Al core levels
far from the junction and is typically an
interface contribution. $\Delta E_b$, on the other hand, is a bulk
contribution and is given by the difference between the binding energies
of the same core levels with respect to the valence band maximum 
and $E_F$ in the
semiconductor and in the metal, respectively: $\Delta E_b = (E_{VBM}^{GaN}-E_{1s}^{Ga})
-(E_{Fermi}^{Al}-E_{1s}^{Al})$. It has been shown that this procedure
is exactly equivalent to the one based on macroscopic average potentials,
traditionally used in  pseudopotential methods\cite{maria}. 

Within this approach
\cite{maria}, the bulk term is the difference between the valence band edge in
the semiconductor and the Fermi level in the metal, each measured with respect
to the average electrostatic potential of the corresponding crystal; the
interface contribution is the electrostatic potential line--up across the
junction.  
However, evaluation of the barrier height in the
presence of electric fields is not so straightforward 
and deserves an appropriate
discussion. In fact, whenever an electric field is present within the
semiconductor side, the core level binding energies (or equivalently the
electrostatic potential) depends on the $z$ position of the atom, so that the
usual procedure given by Eq. \ref{sbh}
becomes ill--defined.

 We therefore considered a linear extrapolation of the
core level binding energies (or of the macroscopic average of the
electrostatic potential) and take the reference energies used in  Eq.
\ref{sbh} as the intercepts of this line with  the two interface planes
(defined in the next paragraph).
We then added the binding energy of the semiconductor side
and obtained two different values, whose distance from $E_F$ in
the superlattice give directly the SBHs. In this way, we
implicitly assume that  $E_F$ in the metal side coincides with the
Fermi level of the superlattice. 
This procedure is exactly equivalent  
to that based on the PDOS
on the different atomic sites, where the SBH is given by the difference of the
valence band maximum on an inner semiconducting atom (where bulk conditions
are recovered) and the Fermi level in the supercell; previous studies
\cite{slv} show
that the results agree with those obtained from Eq.\ref{sbh} within 0.1-0.2
eV.
 
It is now  necessary to
define the two A and B--type interface planes. To this end, we observe in
Fig.\ref{macro_charge}, showing the planar macroscopic average of the valence
charge density, 
that the two interface dipoles - giving rise to
the potential lineup - are centered on the N atom in the B--type interface,
 and half-way between Al and N in the A--type interface.
These planes are (arbitrarily) defined as the interface planes.
We kept this choice  for all our
structures since the dipole locations did not change in the different
systems. Moreover, we can give an
estimate of the uncertainty related to this arbitrary choice,   by
considering interface
planes that coincide with the extremes of the interface
region, namely the interface Al and N  atoms. This leads to an error of
the order of a few hundredths of an eV ($<$ 0.06 eV), so that our overall 
numerical uncertainty on the final SBH values adds up
to 0.15 eV (including also errors in 
 core level binding energies,  
macroscopic potential evaluations, and the accuracy of determining
$E_F$).

Now, since the supercell approach employed introduces artificially two
different interfaces (A and B in Fig.\ref{cell}) and therefore spurious 
boundary conditions, it is worthwhile exploring whether this unrealistic model is 
reliable for the SBH evaluation or if instead it gives an SBH dependent on the  
boundary conditions. To
this end, we considered a larger w-GaN/Al
system (consisting of 17 GaN layers with 9 N, 8 Ga and 6 Al), with exactly the
same interface configuration ({\em i.e.} the same interplanar distances of the
relevant atomic sites in proximity to the junction) as  those 
 previously considered. As shown in Fig. \ref{check_sbh}, we find that 
the electric
field is different (or, equivalently, the potential has a larger slope in the
smaller system), but the SBH remains constant. This clearly shows that: ({\em
i}) the conserved property of the system is the SBH;  electric fields
modify accordingly so as to adjust to boundary conditions,
 keeping constant the difference of the SBH at the
two inequivalent interfaces and  ({\em ii}) our procedure in
estimating the potential line--up is correct.

Let us now discuss the estimated values of the SBH in the different systems,
reported in Table \ref{sbh_tab}. Quasi--particle effects and spin--orbit
coupling have been neglected.   
First of all, note, within our procedure to estimate the SBH, that the
difference between the two potential line--ups at the inequivalent
interfaces ($\Delta \Phi^{A-B}$)  is given by
the electric field times the thickness of the semiconductor region; 
this implies that for an infinite thickness, 
the electric field  would vanish so as to maintain 
$\Delta \Phi^{A-B}$ finite. This is consistent with the 
different charge readjustment occurring at each interface so that the Schottky 
barrier height associated with each isolated interface is a well defined quantity, 
not dependent on boundary conditions\cite{ruini}.

Within the
numerical error estimated above, we observe that  all the values are  
in the range
1.5-2.0 eV.
In particular,  due to the electric fields, 
 the A-type interface has a barrier which is
generally higher than the one at the B interface in the relaxed system, while
the situation is reversed in the ideal structures, consistent with the
discussion in Sect.~\ref{elecfield}.
However, the difference is only about 0.2 eV
since the structural relaxations involved
 are relatively small.

A comparison between the Schottky barriers found here
 and the  barrier value in  
 cubic [001]
ordered z-GaN/Al (1.51 eV\cite{slv}) 
is not straightforward since the interface morphology is very different.
Neverthless, if we compare  bond lengths and 
interface geometry we find stronger similarities with the B-type relaxed
interface, whose SBH  comes out to be  very close to the
one found in the [001] orientation.

As for the comparison with available experimental values, we note that
photoemission measurements performed on Al films deposited on clean 1x1 [0001] GaN
surfaces give a SBH of about 2 eV, which is fully consistent with our
{\em ab--initio} results. This confirms the strong rectifying character of the
GaN/Al contact.

Finally, in order to disentangle the {\em structural} from the
{\em  chemical} contribution to the
 SBH, let us discuss the values for the ideal systems reported in Table
\ref{ideal} together with the corresponding electric fields. 
 First of all, the SBH for the (GaN/Y)$^{id}$ (Y = Ga, Al)
 interfaces are generally smaller
 (from 1.7 to 2.0 eV)
 than those for the  (AlN/Y)$^{id}$ (from 2.4  to 2.8 eV); the rectifying
 behavior of the contact is therefore stronger for the more ionic AlN than for
  GaN. On the other hand, if we keep  the semiconductor
 region fixed and change
 the metallic side of the junction, we note slightly 
smaller modifications of the
 barrier (within 0.3 eV). 
Further, we find,  for Al contacts, that the difference between the
SBH at the two interfaces is smaller than in the Ga junctions: this
is consistent with the smaller electric fields found in the Al case
and with the more effective screening properties of Al discussed above.

Let us now come to the effect of the metallic intralayer 
on the final barrier. 
Any difference in SBH of the XN/X junctions with and without the intralayer has 
   to be ascribed to the interface
 term $\Delta b$ of the potential line--up, since the bulk term $\Delta E_b$ 
 is the
 same. The overall effect is 
 quite small (at most 0.2 eV) and confirms that the change of the
 metallic atom is not so important for the final SBH value, if we keep
the interface geometry constant.

It is interesting to check the potential line--up
transitivity rule \cite{maria} for these ideal systems  in the presence of the
two inequivalent interfaces. In our case, each interface (of A or B--type)
 can be thought of being
composed of different stacked systems:

(XN/Y)$_A^{id}$= (XN/X)$_A^{id}$ + (X/YN)$_B^{id}$ + (YN/Y)$_A^{id}$

(XN/Y)$_B^{id}$= (XN/X)$_B^{id}$ + (X/YN)$_A^{id}$ + (YN/Y)$_B^{id}$

\noindent The transitivity rule is therefore expressed as:

$\Phi^A(XN/Y) = \Phi^A(XN/X) - \Phi^B(YN/X)+ \Phi^A(YN/Y)$

For example, in our case
$\Phi^A(AlN/Ga)$ = 2.42 eV and $\Phi^B(AlN/Ga)$ = 2.84 eV. From the
transitivity rule, we would get in the first case:

$\Phi^A(AlN/Al) - \Phi^B(GaN/Al) + \Phi^A(GaN/Ga)$ =
2.47 -1.80 + 1.67 = 2.34 eV, 

whereas in the second case:

$\Phi^B(AlN/Al) - \Phi^A(GaN/Al) + \Phi^B(GaN/Ga)$ =
2.55 - 1.69 + 1.95 = 2.81 eV.

The agreement with the calculated    $\Phi^A(AlN/Ga)$  and $\Phi^B(AlN/Ga)$ values
is excellent (within the numerical accuracy), so that the transitivity rule  gives
reliable results. Furthermore, this is an additional test of the method proposed to
evaluate the SBHs.

\section{Conclusions}

We have performed {\em ab--initio} FLAPW calculations for the GaN/Al
interface, considering both [111] ordered zincblende and [0001] wurtzite based
systems. Our calculations  focused on the effects that  structural
modifications have on total energies, electric fields and SBH.
We have shown that the structural optimization in the interface region  is, as
expected, very important;  for example, relaxation of the interface Al
positions can even reverse the polarity of the piezoelectric fields due to the
inequivalency of the A and B--type  interfaces.
The final value of the electric field is the result of a complicated interplay
between  boundary conditions, charge rearrangement at the
two junctions and screening effects and cannot be simply justified on the
basis of electronegativity arguments or bond morphology at the
interface.  

On the other hand, the SBH at fixed geometry
is independent of boundary conditions.
However, we have identified some leading mechanisms 
(based on simple ionicity arguments)
 in establishing the final electric fields 
 in the case of unrelaxed systems, due to different chemical
species in the nitride or metallic side of the junctions.
Our procedure to estimate the SBH in the
presence of electric fields is found to give  reliable
results that were  tested by
increasing the unit cell dimensions. 
We have shown that the value of the SBH is not greatly affected by the
presence of the piezoelectric fields of whatever polarity; this 
can only lead to changes
in the SBH of a few tenths of an eV.  Good
agreement with available experimental data is also found.
Finally, the transitivity rule was tested in the case of ideal systems
for both A and B interface types and 
provided SBH values in excellent agreement with the calculated values;
they showed the consistency of our calculations.

\begin{center}
{\bf ACKNOWLEDGEMENTS}
\end{center}

We acknowledge useful discussions with Dr. F. Bernardini and
Dr. P. Ruggerone. Work  supported by the U.S. National Science
Foundation (through the Materials Research Center
at Northwestern University).

   \begin{table}
\caption{Interface bond lengths for the
relevant atomic species (in   \AA) in the fully relaxed systems.}
\vspace{0.2cm}
\begin{tabular}{|c|cc|cc|}
 & \multicolumn{2}{c|}{(w-GaN/Al)$^{rel}$}
&\multicolumn{2}{c|}{(z-GaN/Al)$^{rel}$} \\ & A & B & A & B \\ \hline \hline
$d_{GaN}$ & 1.94 &  1.95 & 1.94 & 1.94\\
$d_{AlN}$ & 1.89 & 1.93 & 1.89 & 1.93 \\
\end{tabular}
\vspace{0.5cm}
\label{distan}
\end{table}

\begin{table}
\caption{Difference between the total energy (in meV/atom)
of the reference system 
(w-GaN[0001]/Al)$^{rel}$ and that of the other interfaces, divided by the
number of atoms.
The estimated error is about 2 meV/atom.}
\vspace{0.2cm}
\begin{tabular}{|c|c|c|c|c|}
 (w-GaN/Al)$^{rel}$ &  (z-GaN/Al)$^{rel}$
&  (w-GaN/Al)$^{pr}$ &  (w-GaN/Al)$^{id}$& (z-GaN/Al)$^{id}$ \\ \hline
0  & 10 & 6 & 55 & 59 \\
\end{tabular}
\vspace{0.5cm}
\label{energy}
\end{table}

\begin{table}
\caption{Electric fields in mV/\AA. The positive sign is relative  to the core
levels being deeper in going from the  B to A interface.
The estimated error is about 0.5 mV/\AA.}
\vspace{0.2cm}
\begin{tabular}{|c|c|c|c|c|}
(w-GaN/Al)$^{rel}$ & (z-GaN/Al)$^{rel}$ & (w-GaN/Al)$^{pr}$
&(w-GaN/Al)$^{id}$ &
(z-GaN/Al)$^{id}$ \\ \hline
23 & 14 & 31 & -5 & -7 \\
\end{tabular}
\vspace{0.5cm}
\label{table_cori}
\end{table}

\begin{table}
\caption{Schottky barrier heights (in eV) at the two inequivalent A and B
type interfaces. The estimated error is about 0.15 eV.}
\vspace{0.2cm}
\begin{tabular}{|c|c|c|c|c|c|}
&(w-GaN/Al)$^{rel}$ & (z-GaN/Al)$^{rel}$ & (w-GaN/Al)$^{pr}$ &
(w-GaN/Al)$^{id}$ & (z-GaN/Al)$^{id}$\\ \hline \hline
$\Phi^A$ & 2.07 & 1.98 & 1.99 & 1.84&  1.69\\ \hline
$\Phi^B$ & 1.69 & 1.76 & 1.48 & 1.93& 1.80\\
\end{tabular}
\label{sbh_tab}
\end{table}

\begin{table}
\caption{Electric fields (in mV/Angstrom)
and Schottky barrier heights (in eV) at the two inequivalent A and B
type interfaces for the unrelaxed systems.}
\vspace{0.2cm}
\begin{tabular}{|c|c|c|c|}
& E & $\Phi^A$ & $\Phi^B$ \\ \hline \hline
(z-GaN/Al)$^{id}$ & -7 & 1.69 & 1.80 \\
(z-GaN/Ga)$^{id}$ & -17 & 1.67 & 1.95 \\
(z-GaN/Al$^{int}$/Ga)$^{id}$ & -5 & 1.91 & 2.00 \\
(z-AlN/Ga)$^{id}$ & -25 & 2.42 & 2.84 \\
(z-AlN/Al)$^{id}$ & -5 & 2.47 & 2.55 \\
 (z-AlN/Ga$^{int}$/Al)$^{id}$ & -18 & 2.40 & 2.70 \\
\end{tabular}
\label{ideal}
\end{table}

\begin{figure}
\caption{Interface geometry at the inequivalent A and B  junctions.}
\label{cell}
\end{figure}

\begin{figure}
\caption{Projected density of states (states/eV-atom) on the interface Ga (panels (a) and
(d)), N  (panels (b) and
(e)) and Al  atoms (panels (c) and
(f))  in the A and B--type (w-GaN/Al)$^{rel}$ interface, respectively.
}
\label{fig_dos}
\end{figure}

\begin{figure}
\caption{N 1s core levels (in eV) vs coordinate along the growth axis (in \AA)
for the ideal (filled circles), partially relaxed (diamonds) and relaxed (grey
squares) GaN/Al interfaces.
Core levels have been arbitrarily shifted so that the central core level in each
of the three systems
coincides with the zero of the energy scale.}
\label{cori}
\end{figure}

\begin{figure}
\caption{Linear fit of the N 1s core levels (in eV) vs growth axis (in \AA)
in the nitride bulk
region. The slope gives the electric field.}
\label{gianni}
\end{figure}

\begin{figure}
\caption{Difference between the macroscopic average of the 
valence charge densities in the
(z-GaN/Ga)$^{id}$ and (z-GaN/Al$^{int}$/Ga)$^{id}$ interfaces 
along the growth axis (solid line). Also shown is its integral (dashed line).}
\label{sandro}
\end{figure}

\begin{figure}
\caption{Macroscopic average of the valence charge density in the
(w-GaN/Al)$^{rel}$ interface along the growth axis.}
\label{macro_charge}
\end{figure}

\begin{figure}
\caption{Macroscopic average of the electrostatic potential (in eV) in the
semiconductor side
of the (w-GaN/Al)$^{rel}$ interface along the growth direction: comparison
between the two cells containing 13 (solid) and 17 (dashed) GaN layers.}
\label{check_sbh}
\end{figure}


\begin{thebibliography}{100}
\bibitem{consist} This is consistent with the observed
possibility of epitaxially
growing cubic instead of hexagonal GaN under specific conditions.
\end{thebibliography}
 
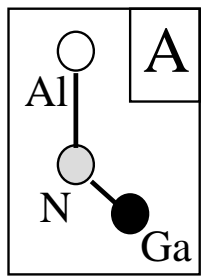
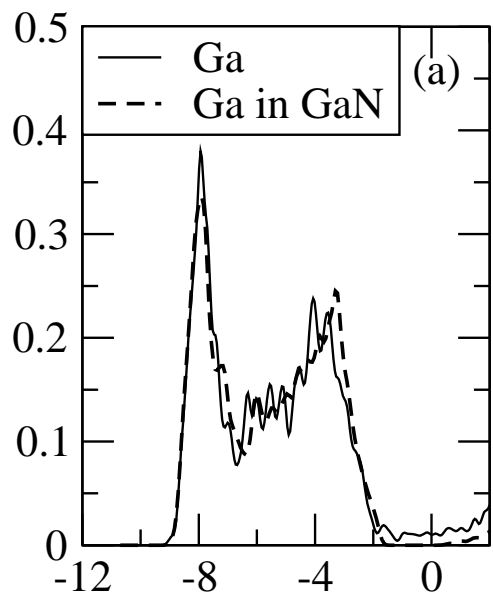
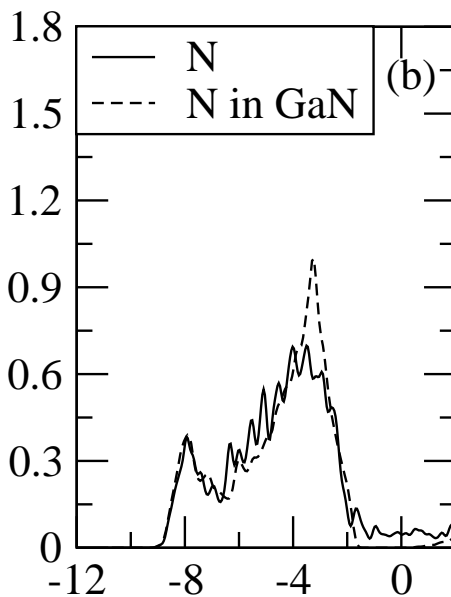
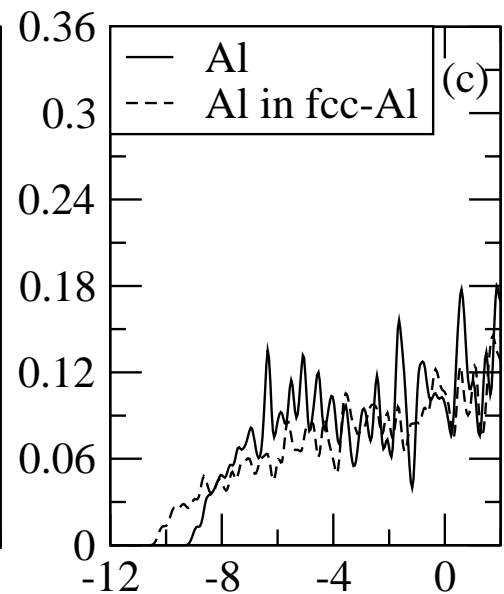
\end{document}